\documentclass{iau}
\usepackage{graphicx}

\title[Molecular emission from SNRs] %% give here short title %%
{The molecular emission from old supernova remnants}

\author[A. Gusdorf, R. G\"usten, S. Anderl, T. Hezareh, \& H. Wiesemeyer]   %% give here short author list %%
{A. Gusdorf, $^1$, R. G\"usten$^2$, S. Anderl$^3$, T. Hezareh$^2$,\\ \& H. Wiesemeyer$^2$}

\affiliation{$^1$LERMA, UMR 8112 du CNRS, Observatoire de Paris, \'Ecole Normale Sup\'erieure, \\24 rue Lhomond, F75231 Paris Cedex 05, France, email: {\tt antoine.gusdorf@lra.ens.fr} \\[\affilskip]
$^2$Max Planck Institut f\"ur Radioastronomie, Auf dem H\"ugel 69, 53121 Bonn, Germany \\
$^3$Argelander Institut f\"ur Astronomie, Universit\"at Bonn, Auf dem H\"ugel 71, 53121 Bonn, Germany}

\pubyear{2013}
\volume{296}  %% insert here IAU Symposium No.
\pagerange{119--126}
\jname{Supernova environmental impacts}
\editors{A. Ray \& D. McCray, eds.}
\begin{document}

\maketitle

\begin{abstract}
Supernovae constitute a critical source of energy input to the interstellar medium (ISM). In this short review, we focus on their latest phase of evolution, the supernova remnants (SNRs). We present observations of three old SNRs that have reached the phase where they interact with the ambient interstellar medium: W28, IC443, and 3C391. We show that such objects make up clean laboratories to constrain the physical and chemical processes at work in molecular shock environments. Our studies subsequently allow us to quantify the impact of SNRs on their environment in terms of mass, momentum, and energy dissipation. In turn, their contribution to the energy balance of galaxies can be assessed. Their potential to trigger a further generation of star formation can also be investigated. Finally, our studies provide strong support for the interpretation of $\gamma$-ray emission in SNRs, a crucial step to answer questions related to cosmic rays population and acceleration.
\end{abstract}

\firstsection 
\section{The astrophysical importance of supernova remnants}

The life of massive stars ends with a supernova explosion, characterized by an important redistribution of energy towards the interstellar medium (ISM). After a free expansion phase (ending when the swept-up mass reaches that of the envelope), and an adiabatic phase (where the energy dissipation is due to expansion), the supernova-driven shocks start radiating energy (\cite{Woltjer72}), initially at observable optical and ultraviolet wavelengths from what has become supernova remnants (SNRs; e.g.\cite{Weiler88}). When SNRs encounter molecular clouds, they drive slower shock waves that compress, accelerate and heat the molecular material and result in strong infrared and sub-millimeter line emission (e.g. \cite{Neufeld07}, \cite{Frail98}). 

\begin{figure} [h]
%\vspace*{-2.0 cm}
\begin{center}
 \includegraphics[width=0.7\textwidth]{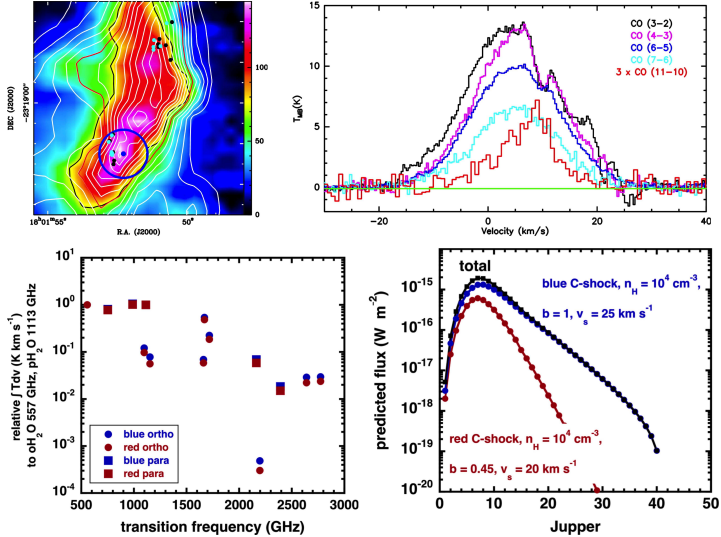} 
% \vspace*{-1.0 cm}
 \caption{CO observations of W28, as presented in \cite{Gusdorf12}. \textit{Top, left:} the field covered by our maps of the CO (6--5) and (3--2) transitions (colours, white contours). The small dots are the OH masers (\cite{Claussen97} and \cite{Hoffman05}). Our shock analysis was made on the position at the center of the circle, that indicates the beam of our SOFIA observations. \textit{Top, right:} the CO spectra (obtained with APEX, (3--2), (4--3), (6--5) and (7--6), and SOFIA, (11--10)), convolved to the SOFIA resolution, were extracted from this position. They were combined with H$_2$ \textit{Spitzer} observations (\cite{Neufeld07}) to constrain our shock models. \textit{Bottom, left:} the models can subsequently be used to make water emission predictions. \textit{Bottom, right:} they can also be used to predict the emission of all of the CO transitions.}
   \label{fig1}
\end{center}
\end{figure}

These relatively slow molecular shocks are found to be very similar to those observed in the jets and outflows associated with star formation (e.g. \cite{Gusdorf11}). However, contrary to their star-formation counterparts, SNR shocks are not expected to be contaminated by the possible UV radiation from the proto-star. Additionally, spectral lines observed in SNR shock regions do not either show any envelope or infall component (see for instance the upper right panel in Fig.\,\ref{fig1}) that would make their interpretation complicated. SNRs hence serve as clean laboratories to study the physical and chemical mechanisms that operate in shock environments.

In a further step, studying SNRs interacting with the ISM is a powerful tool to quantify their contribution to the energy balance of galaxies, through the observation of the numerous CO transitions recently allowed by the \textit{Herschel} telescope, and presented in flux diagrams, the so-called CO ladders (e.g. \cite{Haileydunsheath12}). In certain cases, their potential to trigger star formation can be investigated, see \cite{Xu11}.

Finally, the molecular emission from SNRs provide a valuable support for the study of cosmic rays population (CRs, hadronic or leptonic) and acceleration. Because of their large energy budget, it has indeed long been argued that SNRs are the primary sites for accelerating CRs (e.g., \cite{Blandford87}). Observing the $\gamma$-ray emission in SNRs environments is a way to constrain the CRs population, and acceleration mechanisms. Indeed, this radiation mainly results from three processes: $\pi^0$ decay from the interaction of the hadronic component of the CRs population with the ambient medium, Bremsstrahlung emission from the leptonic component on the ambient ISM, and inverse Compton scattering of the leptonic component on the ambient radiation field, (e.g. \cite{Frail11} for a short review). Although the situation is a bit more complex for old SNRs (e.g., \cite{Gabici09}, \cite{Bykov00}), whose shocks no longer accelerate CRs, an accurate knowledge of the ambient medium (density, mass of the shocked/non shocked gas, local magnetic field strength, local radiation field) remains crucial to assess the contribution from each of these three processes to the very high energy spectra observed in SNRs.

We demonstrate how the observation of the molecular emission from three old SNRs, W28, 3C391, and IC443, allow to address these diverse astrophysical topics.

\section{Shock modelling in W28F}

The upper panels of Figure\,\ref{fig1} summarize the results presented in \cite{Gusdorf12}, dedicated to observations and models of shocked regions in the W28 SNR. By analysing the CO (from APEX and SOFIA telescopes) and H$_2$ (from the \textit{Spitzer} telescope) emission, we have constrained shock parameters in the beam of our observations. For this position, we have extracted two CO integrated intensity diagrams, one per velocity range (blue lobe, -30 to $\sim$10~km~s$^{-1}$, and red lobe, $\sim$10 to 40~km~s$^{-1}$). We have compared the corresponding high-$J$ CO emission, that most unambiguously traces the shocked material, to a grid of unidimensional shock models. We hence have fitted each velocity component by a single, C-type shock wave arising from a 25$''$ diameter emission region, with the respective parameters: pre-shock density $n_{\rm H}$ = 10$^4$~cm$^{-3}$, magnetic field strength perpendicular to the shock front $B$ = 100 and 45 $\mu$G, and shock velocities $v_{\rm s}$ = 20 and 25 km s$^{-1}$. We also checked that our models provide reasonable fits for H$_2$ excitation diagrams (\textit{Spitzer} observations by \cite{Neufeld07}). Our final results are compatible with independent studies of the region in terms of age (\cite{Giuliani10}), magnetic field measurements (\cite{Claussen97}, \cite{Hoffman05}), and densities required for the excitation of observed OH masers (\cite{Lockett99}).

In a forthcoming publication (Gusdorf et al., in prep), we will go one step further in the interpretation of the results. With the shock model parameters, and assumptions on the size of the emitting region, we can thus infer the shocked mass in the beam of our observations (from 6.2 to 18.5 $M_\odot$ depending on the adopted age, 10$^4$ or $3\times10^4$ years, respectively). Combining our shock model results with an LVG code to calculate the emission from water (\cite{Gusdorf11}), we are also able to predict integrated intensity diagrams for each velocity component, which can be compared for instance to observations made by the \textit{Herschel} telescope, as can be seen on the lower left panel of the Figure\,\ref{fig1}. Additionally, we predict the excitation from all CO lines in our beam (lower right panel of the Figure\,\ref{fig1}), that can be directly compared with \textit{Herschel} observations of the region. Such SNR shocks CO ladders can also directly be compared an help constraining their contribution to galactic ones (e.g. \cite{Haileydunsheath12}).

\section{The cases of 3C391 and IC443}

Forthcoming publications will also be focused on a similar analysis of SNRs 3C391 and IC443 (Gusdorf et al., in prep), based on CO and H$_2$ observations.

\begin{figure}[h]
% \vspace*{-2.0 cm}
\begin{center}
 \includegraphics[width=0.8\textwidth]{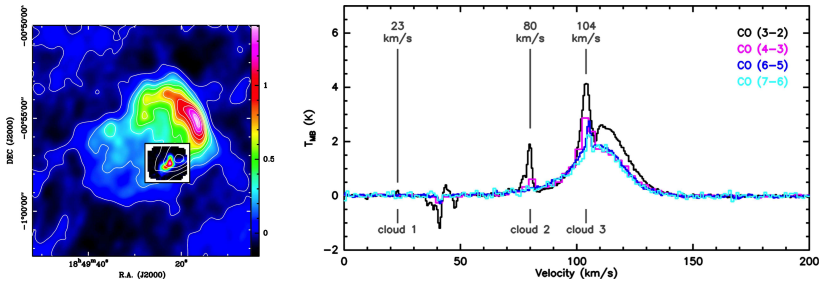} 
% \vspace*{-1.0 cm}
 \caption{\textit{Left:} 325 MHz radio-continuum emission in the 3C391 SNR, as it appeared in \cite{Moffett94} (colours and white contours). The small inset is the field of our CO observations with the APEX telescope, and shows the CO (6--5) emission. \textit{Right:} the spectra in various CO lines observed with the APEX telescope, averaged over the whole observed field.}
   \label{fig2}
\end{center}
\end{figure}

In 3C391, we find that our APEX CO observations directly allow to identify the line-of-sight clouds that lie in the vicinity of the remnant, and constitute potential targets for the cosmic rays interactions. Figure\,\ref{fig2} shows both the field of our observations (left panel), and the averaged CO spectra we extracted (right panel). Based on our CO and $^{13}$CO observations, we were able to infer the mass of the clouds that might be impacted by the cosmic rays accelerated in the SNR, the so-called clouds 2 and 3, respectively 161.3 and 43.7 $M_\odot$ - see for instance \cite{Xu11} for the calculation method.

\begin{figure}[h]
% \vspace*{-2.0 cm}
\begin{center}
 \includegraphics[width=0.5\textwidth]{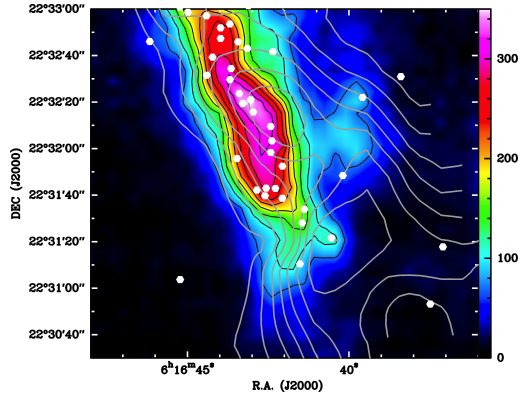} 
% \vspace*{-1.0 cm}
 \caption{The G clump of the IC443 SNR, mapped with APEX in the CO (6--5) transition (in colours and black contours, Gusdorf et al., in prep). The white markers indicate the position of YSO candidates in the region, selected in the 2MASS point source catalog, based on colour selection criterion by \cite{Xu11}. Also shown in the greyscale contour is the CO (1--0) ambient emission at 5 km s$^{-1}$, observed by the IRAM 30m telescope (Hezareh et al., in prep).}
   \label{fig3}
\end{center}
\end{figure}

Finally, in the G clump of the IC443 SNR, our observations (Gusdorf et al., in prep, and Hezareh et al., in prep, and Figure\,\ref{fig3}) confirm the results obtained by \cite{Xu11}. As the YSO candidates in the region are older than the SNR, these authors show that in spite of the correlation between their distribution and the shock structure, star formation has probably not been triggered by the supernova-driven shock wave, but rather by the stellar winds of the massive progenitor of the remnant. Paradoxically, our observations confirm the tight correlation between the star formation and the shocked CO (6--5) gas in the region. On the other hand, our CO (1--0) observations show that the star formation has developed on the edge of a nearby ambient cloud (also see \cite{Lee12}), subject to the stellar winds of the SNR progenitor prior to the passage of the shock wave.

\end{document}